\documentclass[aps,preprint,showpacs,showkeys]{revtex4}
\usepackage{graphicx}
\usepackage{amssymb}
\usepackage{bm}
\usepackage{epsfig}

\begin{document}
\setcounter{page}{1}
\title{The Spread of the Credit Crisis: View from a Stock Correlation Network}
\author{Reginald D. Smith}
\affiliation {Bouchet-Franklin Research Institute \\ PO Box 10051, Rochester, NY 14610}
\email {rsmith@bouchet-franklin.org}


\begin{abstract}
The credit crisis roiling the world's financial markets will likely
take years and entire careers to fully understand and analyze. A
short empirical investigation of the current trends, however,
demonstrates that the losses in certain markets, in this case the US
equity markets, follow a cascade or epidemic flow-like model along
the correlations of various stocks. This phenomenon will be shown by
the graphical display of stock returns across the network and by
the dependence of the stock return on topological measures.
\end{abstract}


\maketitle
\section{Introduction}
The barely covered story of rising foreclosures among the
condominiums of Florida or California in early 2007 was a harbinger
of a much larger collapse in the worldwide financial system. The
increase of foreclosures over the priced in foreclosure risk in
mortgage-backed securities, otherwise deemed high-grade assets,
began the confusion of the value of collateral assets and subsequent
seizing up of credit markets around the globe. The collapse of
several institutions, such as Bear Stearns, Lehman, and Fortis, has
accentuated the level of crisis now facing the world markets.
Previously, loosely regulated titans of finance, such as hedge funds
and private equity groups, have been hit by waves of unprecedented
losses and demands by investors for redemptions, causing them to
sell even more assets or close positions and creating a positive
feedback death spiral.

Though the hardest hit markets are lesser-known markets, such as
commercial paper, the equity markets have become the most widely
known indicators of the ongoing meltdown. In fact, most non-experts
likely use the movements of the equity markets, fallaciously, as a
key gauge of the severity or progress of the crisis. The equity
markets, however, did not originate the crisis nor are they the key
force perpetuating it. In this short paper, the spread of the credit
crisis will be discussed by referring to a correlation network of
stocks in the S\&P 500 and the NASDAQ-100 indices. The fact that the
spread resembles a contagion or cascade, however, may be mainly
superficial given the underlying dynamics are completely different.

\section{Network Construction}
In this paper, a stock correlation network, similar to the
one in Refs.
\cite{econophysics1,econophysics2,econophysics3,econophysics4,econophysics5}, is created.
We start by defining a correlation matrix of returns between two stocks, where the correlation between stocks $i$ and $j$,
$\rho_{ij}$ is defined as

\begin{equation}
\rho_{ij} = \frac{E((X_i - \mu_i)(X_j - \mu_j))}{\sigma_i\sigma_j}
\end{equation}

with $X_i$ and $X_j$ being the log-returns of stocks $i$ and $j$ at a
given time, $\mu_i$ and $\mu_j$ being the mean value of the stock
log-returns over the measured time period, and $\sigma_i$
and$\sigma_j$ being the standard deviations of $i$ and $j$ over the
measured time period. The correlation is taken over the time period August 1, 2007, to October 10, 2008, where each daily value of $X$ is the log-return of
the closing price from the previous day. As Refs.
\cite{econophysics1,econophysics2} demonstrate, however, correlation
is not a distance metric; therefore, we create an adjacency matrix
with weights on the edges matching the distance metric between
stocks, $i$ and $j$. That matrix is defined as

\begin{equation}
d = \sqrt{2(1-\rho_{ij})}
\label{distanceeq}
\end{equation}

\begin{figure}[t!]
    \includegraphics[width=13.0cm]{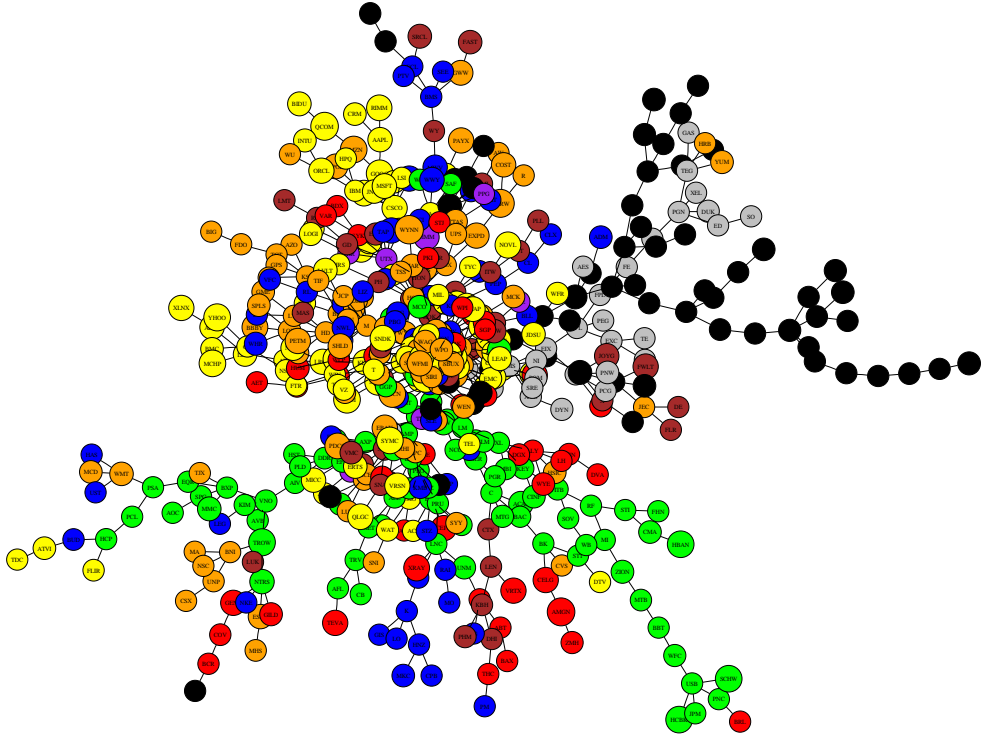}
    \caption{Sectors represented by stocks in the network. Green is for
    finance firms, orange is for service firms, red for healthcare, grey for
    utilities, yellow for technology firms, black for basic materials, purple for conglomerates, blue for consumer goods, and brown for industrial goods. Industry sector breakouts are according to Hoovers\cite{hoovers}.}
    \label{sector}
\end{figure}

Using these distances we finally create a graphical minimal spanning tree
by using the python-graph module, pydot, and Graphviz.
Because over 500 stocks are included, the ticker labels are
relatively small but the central part of the component is dominated
(though not exclusively) by certain finance and service sector stocks, which
are heavily cross-correlated and thus tightly linked with each
other, while the outer branches are more industry specific, including
utilities, basic materials, technology, and some less-central financial stocks. These are the stocks later impacted by the credit crisis (see Fig. \ref{sector}). The average
correlations among stocks both within each category and between
stocks of each category are given in Table \ref{corrtable}.
\begin{table*}[!t] \vspace{1.5ex}
\tiny{
\begin{tabular}{|c|c|c|c|c|c|c|c|c|c|}
\hline &Basic Materials&Conglomerates&Consumer
Goods&Financial&Healthcare&Industrial
Goods&Services&Technology&Utilities\\
\hline
Basic
Materials (61)&0.65&0.68&0.46&0.52&0.46&0.62&0.52&0.58&0.6\\
\hline
Conglomerates (7)&0.68&0.88&0.62&0.69&0.60&0.79&0.7&0.74&0.75\\
\hline
Consumer
Goods (61)&0.46&0.62&0.48&0.53&0.45&0.56&0.52&0.53&0.55\\
\hline
Financial (85)&0.52&0.69&0.53&0.64&0.49&0.63&0.59&0.59&0.6\\
\hline
Healthcare (49)&0.46&0.60&0.45&0.49&0.46&0.53&0.49&0.51&0.54\\
\hline
Industrial Goods (42)&0.62&0.79&0.56&0.63&0.53&0.71&0.63&0.66&0.66\\
\hline
Services (98)&0.52&0.7&0.52&0.59&0.49&0.63&0.59&0.60&0.61\\
\hline
Technology (100)&0.58&0.74&0.53&0.59&0.51&0.66&0.60&0.65&0.63\\
\hline
Utilities (30)&0.60&0.75&0.55&0.60&0.54&0.66&0.61&0.63&0.76\\
\hline
\end{tabular}
}  \caption{Correlations within and across stock categories from
8/1/1007 to 10/10/2008. The number in parentheses after the sector name
in the rows is the number of companies in each category.}
\label{corrtable}
\end{table*}

\begin{figure}[t!]
    \includegraphics[width=13.0cm]{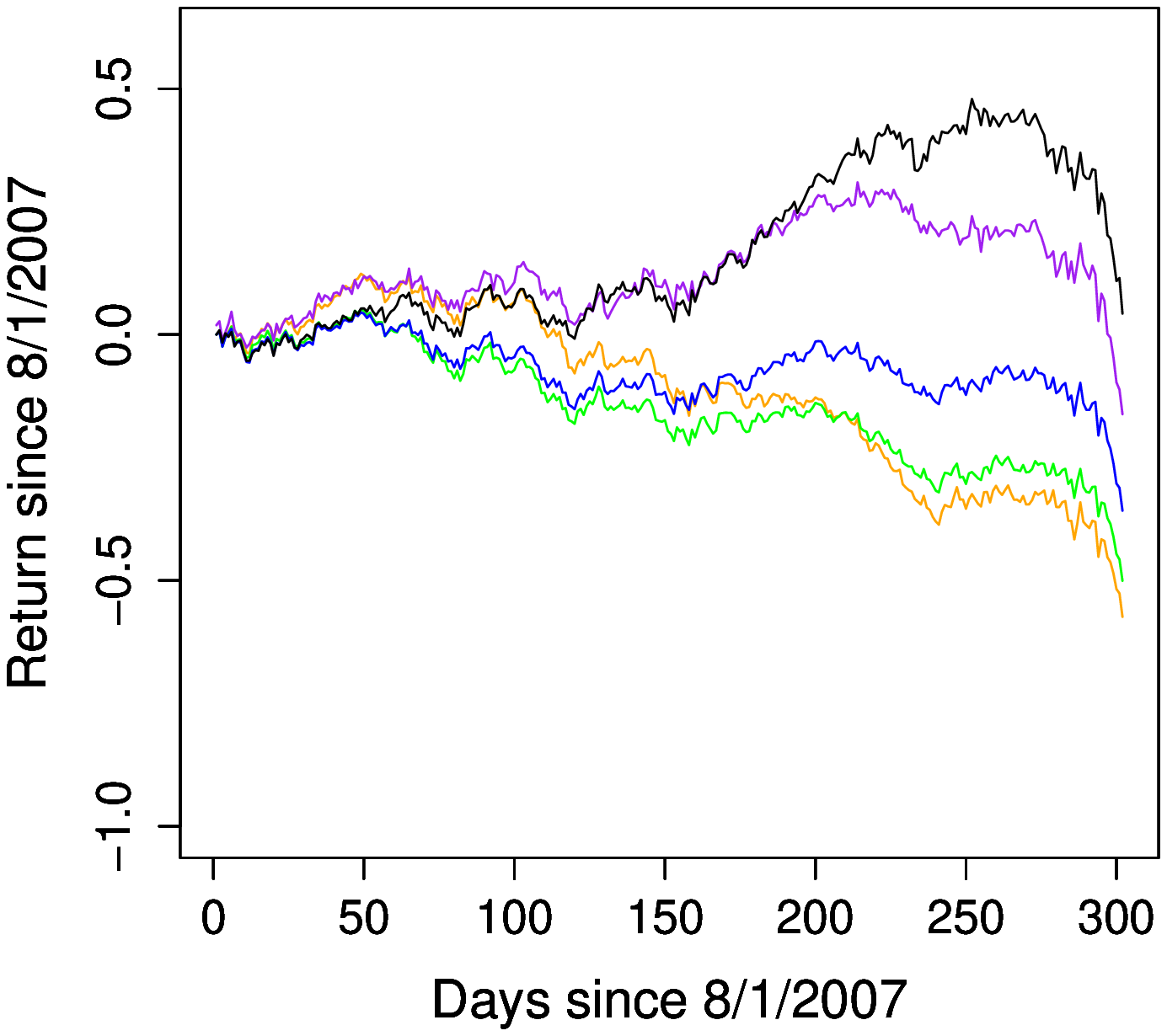}
    \caption{Average returns of stocks from August 1, 2007 by
    distance from the stock with the highest betweenness centrality
    (CBS). Orange is the average return of stocks at distance $0<d\leq 0.4$,
    green at distance $0.4<d\leq0.8$, blue at distance $0.8<d\leq1.2$, purple at distance $1.2<d\leq1.6$, and black at distance $1.6<d\leq2.0$ (the maximum allowed by the metric in Eq. \ref{distanceeq}).}
\label{distanceplot}
\end{figure}

\begin{figure*}[t!]
\tiny{
\centering
\begin{tabular}{cc}
    \includegraphics[height=2.0in, width=2.0in]{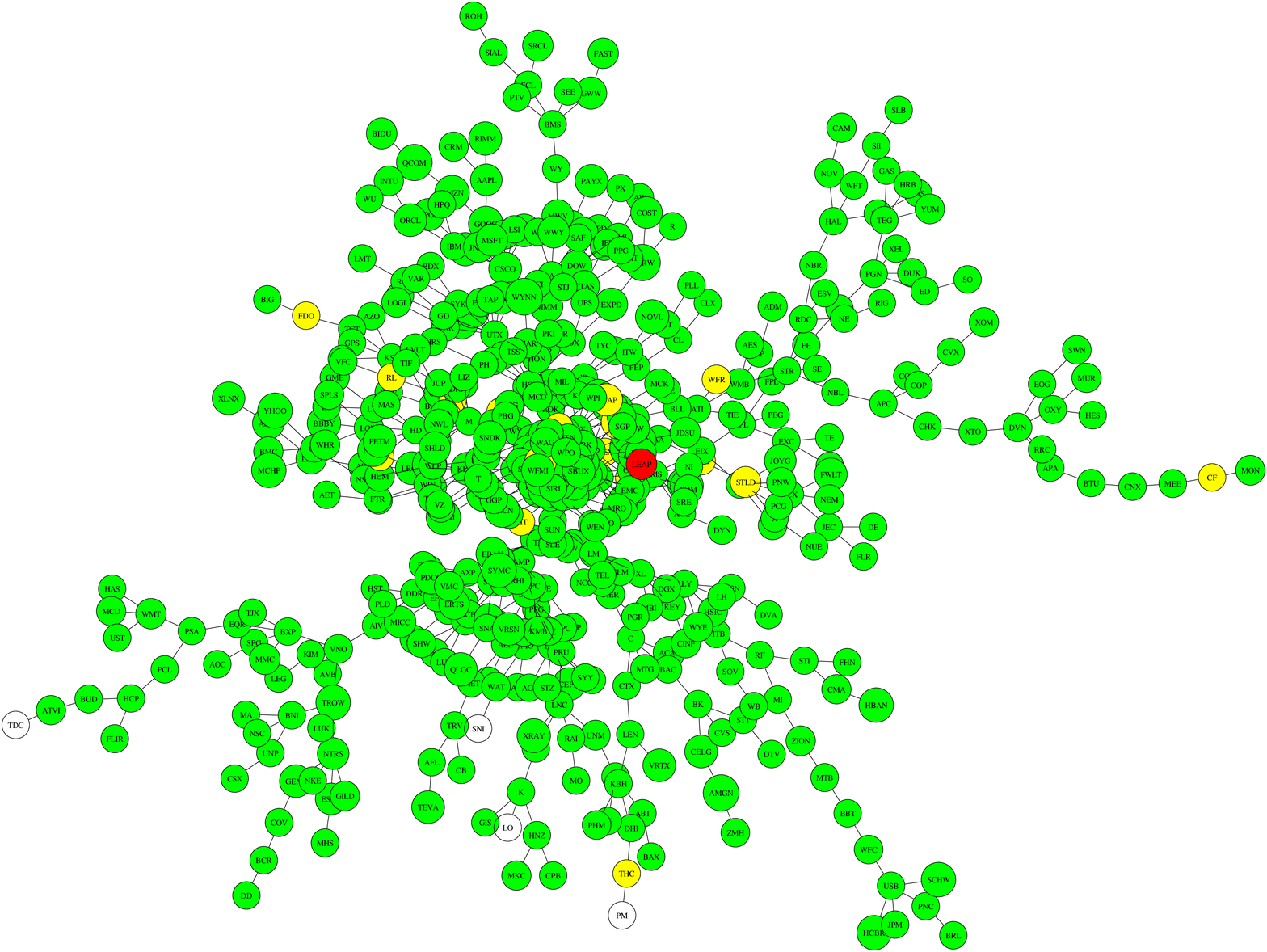}&
    \includegraphics[height=2.0in, width=2.0in]{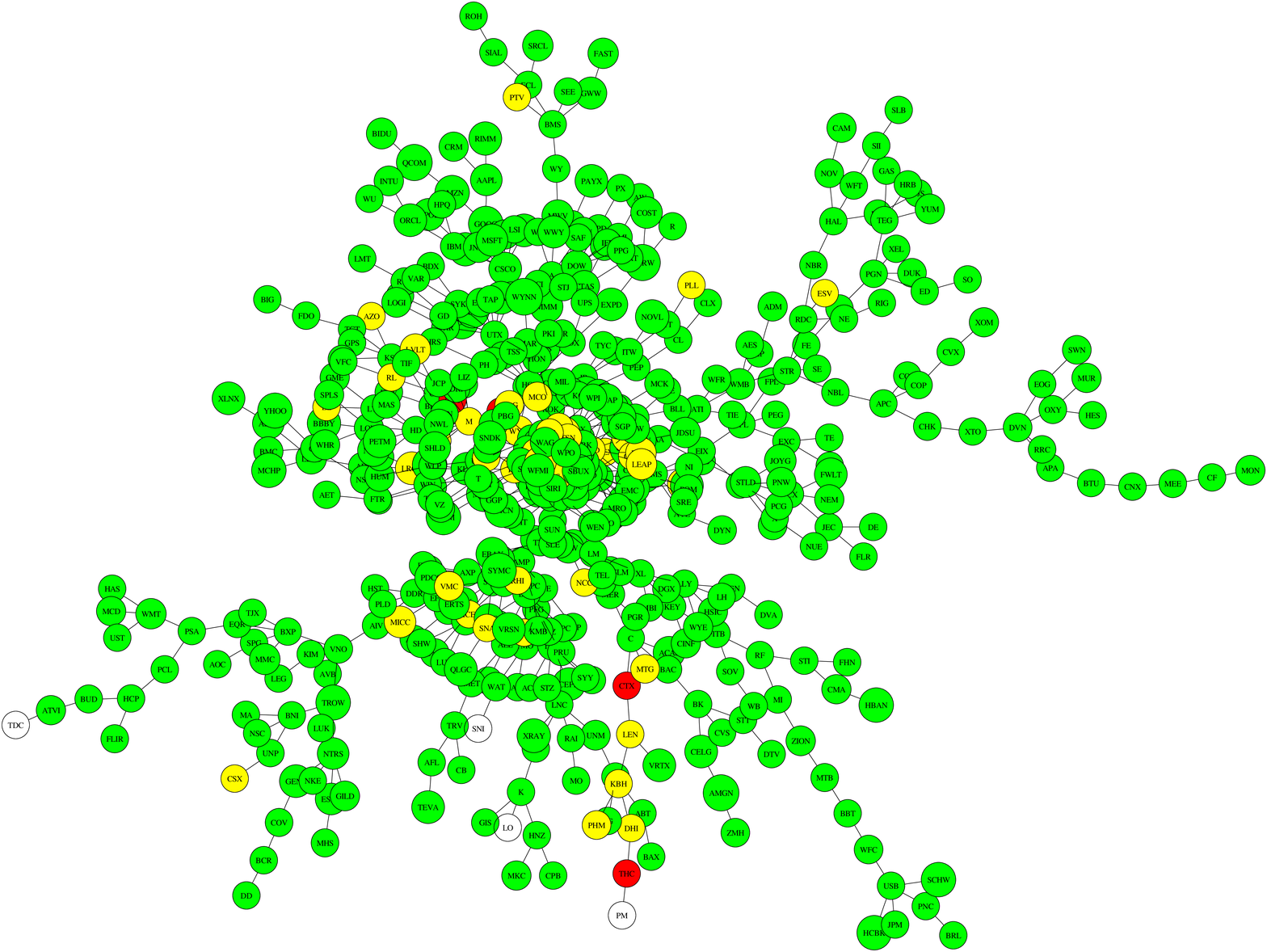}\\
    (a)&(b)\\
    \includegraphics[height=2.0in, width=2.0in]{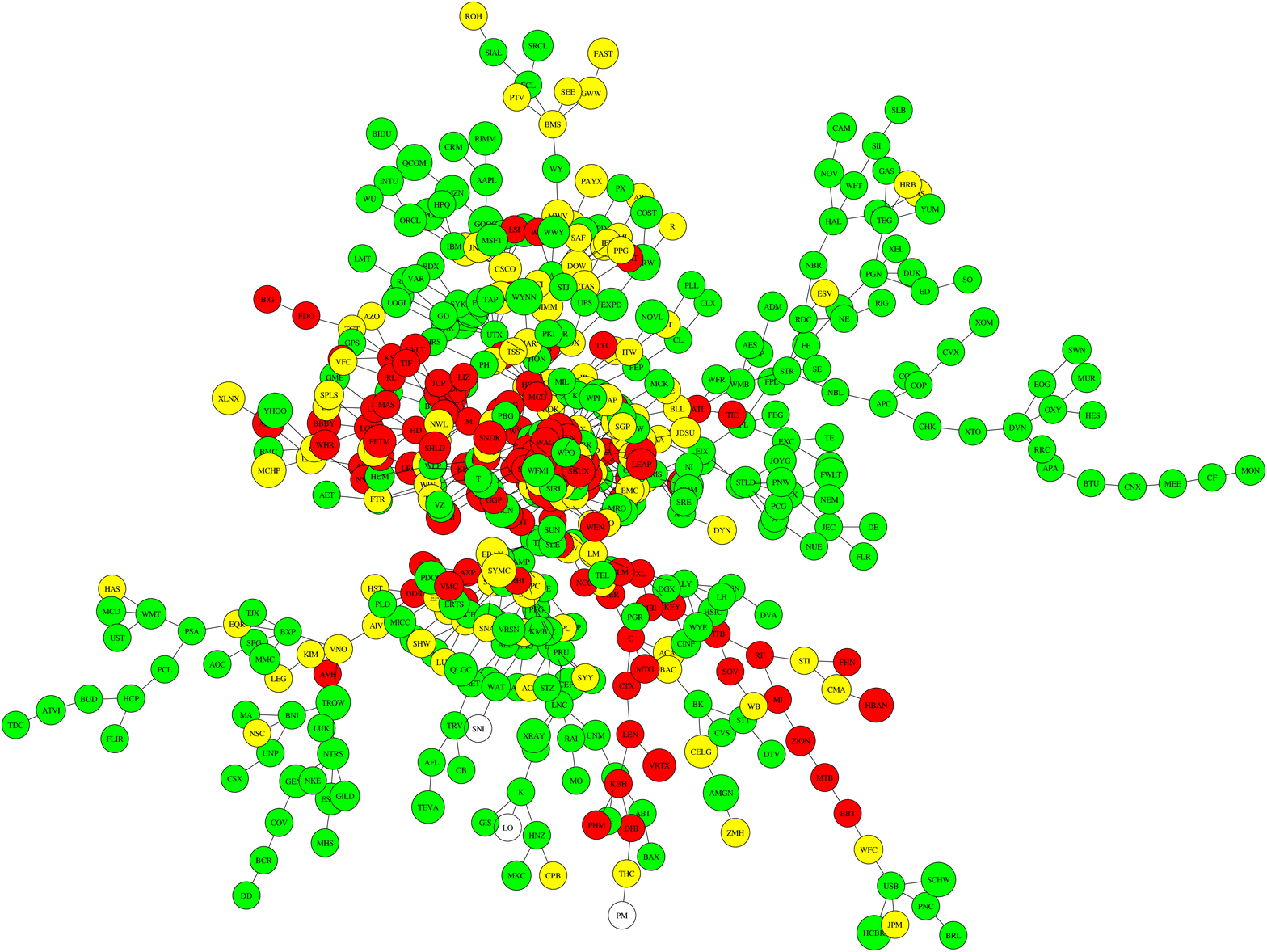}&
    \includegraphics[height=2.0in, width=2.0in]{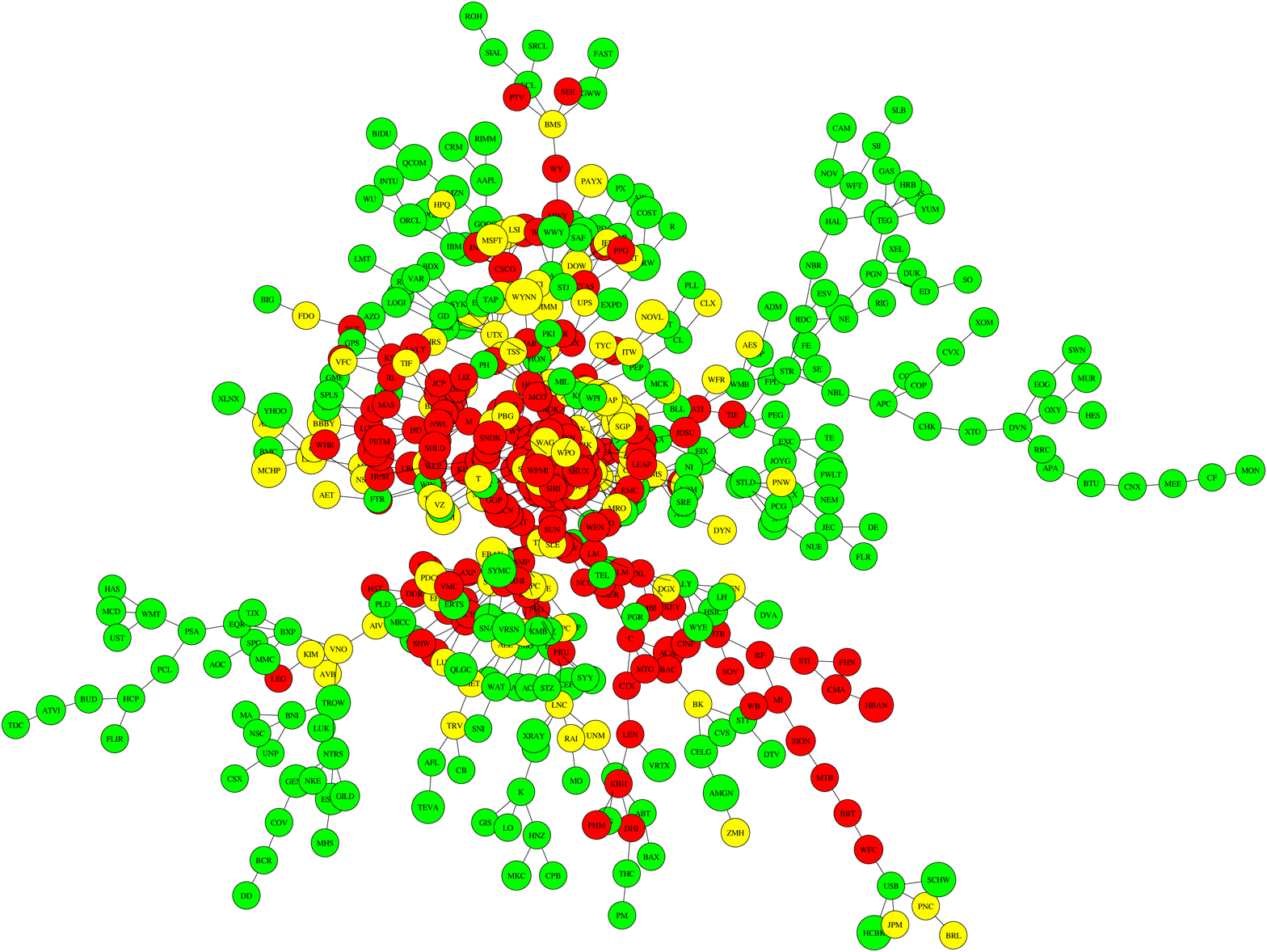}\\
    (c)&(d)\\
\includegraphics[height=2.0in, width=2.0in]{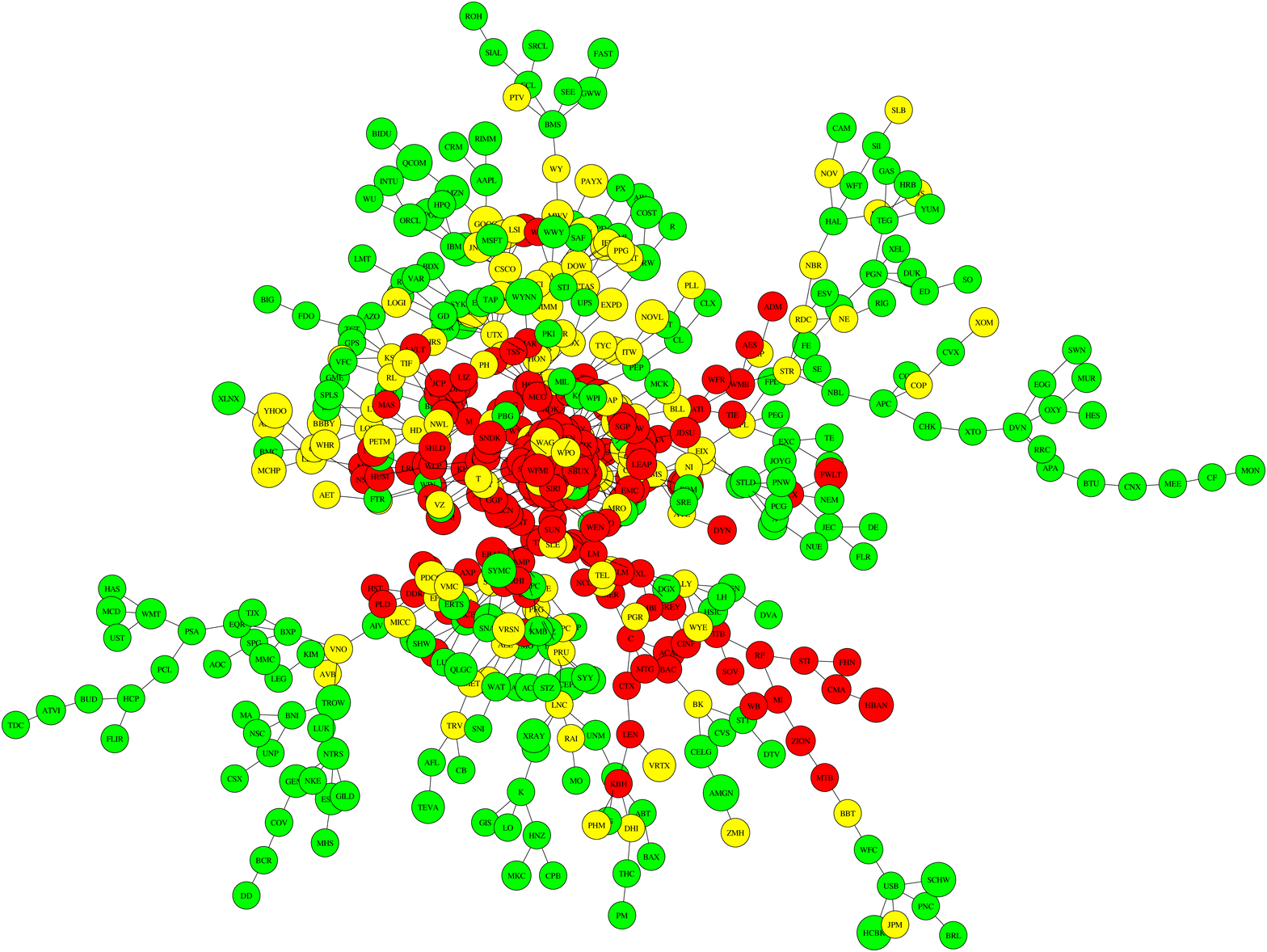}&
    \includegraphics[height=2.0in, width=2.0in]{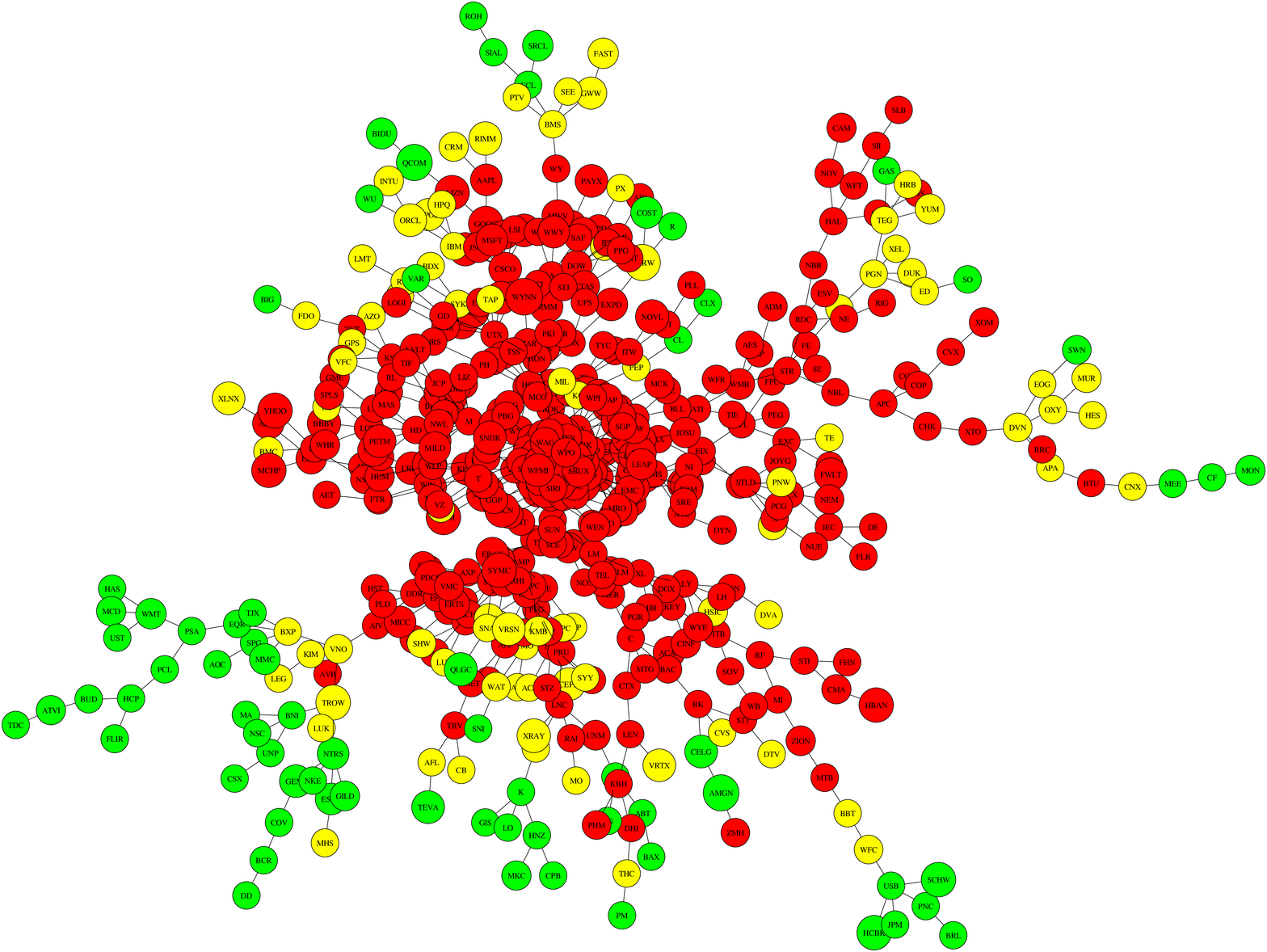}\\
        (e)&(f)\\
\end{tabular}
\caption{Diagrams show the spread of the credit crisis across nodes of the stock correlation
        network for different dates. From the top left, (a)August 10, 2007, when the crisis in mortgage-backed securities first
        began to cause widespread market volatility, (b) September 14, 2007, the collapse of British lender Northern Rock and its bailout by the British
        government, which accentuated the global spread of the crisis, (c) January 17, 2008, turbulence in January 2008 due to the increasing fear
        of instability in the financial sector, (d) March 17, 2008, the collapse of the once venerable Wall Street investment bank Bear Stearns,
        (e) September 15, 2008, the even more destabilizing collapse of Lehman Brothers, and (f) October 10, 2008,
        end of the worst performing week for the Dow Jones Industrial Average in history. Green nodes represent a current
        arithmetic return greater than -10\%. Yellow nodes represent a current return between -10\% and
        -25\%. Red nodes represent a current return less than -25\%. }
\label{spread}
}
\end{figure*}

The stocks in Fig. \ref{spread}, represented as nodes, are colored
according to the following methodology based on the stock return
since August 1, 2007. Events in the figures are taken from the timeline at Ref.
\cite{timeline}. The fall in stock valuations flows outward in the correlation
network from stocks with relatively high centrality in the center to
those on the periphery, which are more industry specific or otherwise
uncorrelated to the core sectors of the stock market. In Fig.
\ref{distanceplot}, this spread is emphasized by showing the average
return among stocks at a distance $d$ from the stock with the
highest betweeenness centrality (here CBS, a major S\&P 500 stock,
and here classified under the services industry), where $d$ is
defined by Eq. \ref{distanceeq}. Here, we see that the greater
the distance from the central part of the network, the more delayed
the decline in valuation. Therefore, the credit crisis spreads among
affected stocks from more centralized nodes to more outer ones as
the news of the extent of the damage to the global economy spreads.

\section{Discussion \& Conclusion}

Using methods of statistical physics and complex networks to
investigate phenomena in stock markets is increasingly common
\cite{chinastock1,chinastock2,koreanstock1,koreanstock2}. The
increasing complexity and globalization of financial markets has led
to many large and sometimes unpredictable effects. In Ref.
\cite{koreanstock3}, the effects of globalization upon the Korea
Stock Exchange were demonstrated by showing the increasing grouping
coefficient of stocks from 1980-2003. The credit crisis, however,
presents a challenge of a whole new magnitude.

As viewed by the wider market, the collapse in stock price
returns began in the financial and services sector of the economy.
Soon it moved across more mainline banks and firms, and more
recently has affected stocks across the board. Though the spread of
the collapse in stocks down the tree resembles an infection or
cascade on a network, such ideas are more appropriately viewed as
analogies or metaphors than explanations. Unlike a disease or
cascading collapse, the stock crash is not being transmitted from
one stock to another. What the collapse reveals is a complex and
collective systemic collapse of the financial system, which spreads
as its extent becomes more recognized and affects the credit or
demand for sectors across the economy.

The spread is carried both by the news of the extent of the crisis
and the fact that similar risky asset bases make the co-movement of certain stocks more likely and thus more highly correlated. In addition, as credit becomes restricted, capital flows formerly relied on as a given begin to disappear, causing financial difficulties in companies and selling of equities (among other assets) to raise
capital. As panic and the extent of the devastation spread, stocks
are punished accordingly. In normal times, the failure of a company
and its stock is not a cause for a systemic crisis. Also, since the
correlation was calculated over an entire year's activity, the stock
prices are correlated because they tend to fall similarly over time.
The correlation shown in this network does not cause the
transmission chain of collapse, but is inextricably tied to it. In
addition, the correlation generally increases with volatility (for
example, see Ref. \cite{finance1}) and negative returns affect volatility
more than positive returns of the same magnitude
\cite{finance2,finance3}, so over time, the correlation has been
increasing among stocks, and the network will likely be more dense
and structured differently due to the steadily increasing market
volatility.

Finally, one should note that this is not an example of the widely cited
`financial contagion' in the press. Financial contagion refers to
the coupling of financial panic across national borders and not
among stocks in an exchange. However, these do illustrate the spread
of the credit crisis and how what was once a problem among home
builders and mortgage finance companies has engulfed the entire
economy.

\end{document}